\documentclass[aps, superscriptaddress,preprintnumbers, twocolumn ]{revtex4-1}

\usepackage{amsmath}
\usepackage{amssymb}
\usepackage{graphicx}
\usepackage{dcolumn}
\usepackage{bm}
\usepackage{color}
\usepackage{multirow}

\begin{document}

\title{The role of face-to-face lecturing in flipped classrooms}

\date{\today}

\author{Heeseung Zoe}
\email{heezoe@dgist.ac.kr}
\affiliation{School of Basic Science, DGIST, Daegu 711-873, Republic of Korea.}
\affiliation{Division of General Studies, UNIST, Ulsan 689-798, Republic of Korea}
\affiliation{Center for Learning and Teaching, UNIST, Ulsan 689-798, Republic of Korea}

\author{Kyujin Kwak}
\affiliation{Department of Physics, School of Natural Science, UNIST, Ulsan 689-798, Republic of Korea}

\author{Yoonjoo Cho}
\affiliation{Division of General Studies, UNIST, Ulsan 689-798, Republic of Korea}
\affiliation{Center for Learning and Teaching, UNIST, Ulsan 689-798, Republic of Korea}

\author{Yun-Young Seo}
\affiliation{Center for Learning and Teaching, UNIST, Ulsan 689-798, Republic of Korea}

\author{Jin-suk Choi}
\affiliation{Division of General Studies, UNIST, Ulsan 689-798, Republic of Korea}

\author{Jin-Hyouk Im}
\affiliation{Center for Learning and Teaching, UNIST, Ulsan 689-798, Republic of Korea}
\affiliation{School of Business Administration, UNIST, Ulsan 689-798, Republic of Korea}

\author{Soogyun Beum}
\affiliation{Center for Learning and Teaching, UNIST, Ulsan 689-798, Republic of Korea}

\author{Hai-Woong Lee}
\affiliation{Division of General Studies, UNIST, Ulsan 689-798, Republic of Korea}
\affiliation{Department of Physics, School of Natural Science, UNIST, Ulsan 689-798, Republic of Korea}

\begin{abstract}
As flipped learning enforces students' self-preparation for the class, in-class unidirectional lectures can be replaced with interactive teaching methods such as peer instruction and group activities.
In this article, we study the role of face-to-face lecturing in introductory physics classes at the university level based on the flipped learning, by analysing student achievements and the degree of student satisfaction for two semesters.
In the first semester, we redesigned a class of 176 students with a reduction in lecture hours and replaced them with pre-class self studies and in-class group problem solving,
and compared it with the traditional lecture-based class of 161 students.
It is confirmed that the redesigned class offered students a more effective way of learning physics than the traditional lecture-based class.
In the second semester, we applied the same strategy of course redesign to two classes but controlled the proportion of face-to-face lectures to the entire class meeting hours systematically to be 1/3 in \textit{Class 1} of 160 students and 1/2  in \textit{Class 2} of 176 students.
It is found that the proportion does not seem to affect the students' achievements in those two classes.
However, the students who have performed well in the midterm and final exams, tended to think that the group problem solving was more helpful than the face-to-face lecture, while the students whose final exam scores are relatively low compared to the midterm exam scores tended to prefer the face-to-face lecture.
\end{abstract}

\maketitle

\section{Introduction}

Recently a popular trend of course redesign is to arrange educational tools for active learning, which increases the interactions between instructors and students in class, rather than unilaterally imparting instructors' knowledge in the form of traditional lectures \citep{hake,mcdermott,potter,meltzer,pollock, west,rudolph,goldberg,sahin}. The course elements which make students engage with their peers in in-class activities are fruitful especially for physics education in the forms of cooperative problem solving \citep{heller1,heller2,bruun,dori} and peer instruction \citep{mazur1,mazur2,smith,miguel}. Even in physics classes with a high enrollment, a relevant strategy for course redesign may overcome the challenges from the massive audiences and resolve problems due to the lack of experience that a young lecturer may have \citep{wieman}.

One of the ways of increasing the interactions in classrooms is `flipped learning' \citep{bergmann1,bergmann2}.
The 2014 Horizon report specifies that `flipped classroom' is one of the most important developments in educational technology for higher education to be implemented in one year or less \citep{horizon}.
According to a recent survey by the Center for Digital Education, 50\% of American faculty members are already using the flipped learning method or plan to by winter 2014 \citep{webinar}.

\begin{figure*}[ht]
\centering
\includegraphics[scale=0.6]{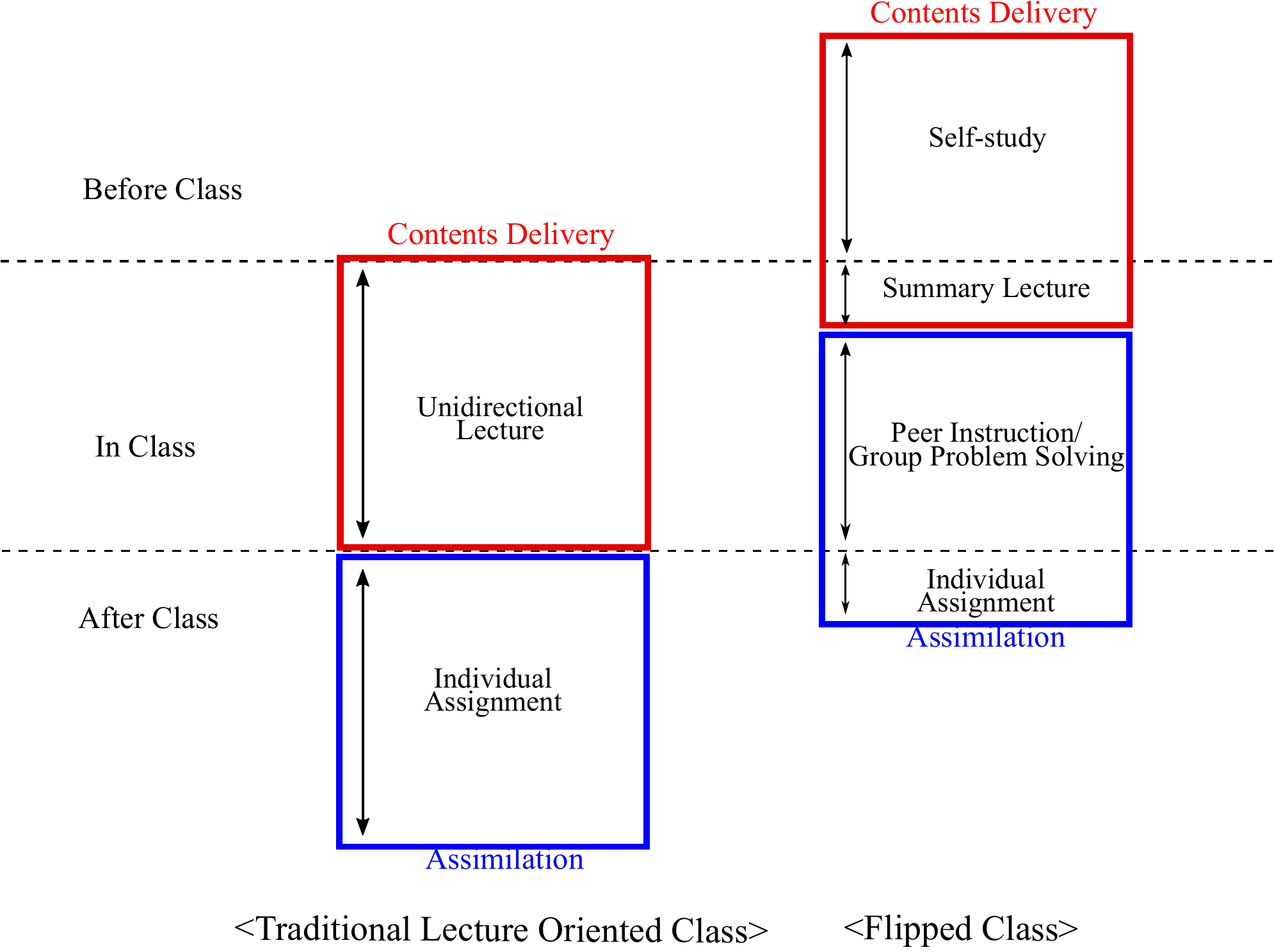}
\caption{ Comparision of the structures of traditional lecture oriented class and flipped class.
 }\label{fig:flipped}
\end{figure*}

As discussed in \citep{bates}, the basic structure of flipped learning is a combination of Just-In-Time-Teaching \citep{novak} and the student engagement components like group problem solving \citep{heller1,heller2} and peer instruction \citep{mazur1,mazur2}.
It is to flip the learning process of the lecture-oriented class as in Fig. \ref{fig:flipped}.
In the traditional lecture-oriented class, instructors do their best to deliver lectures on contents to students as a main in-class activity, and then assimilation is passed unto students e.g. by assigning homework.
However, in flipped classroom, the course contents are posed to students as a form of pre-class self-study in advance, and the main in-class activity is to discuss the contents or solve related problems with instructors and their peers.
In the framework of flipped learning, instructors may arrange various interactive in-class activities for student engagement and assessment.

It is a challenge for those adopting flipped learning strategy to produce and develop pre-class materials like movie clips, lecture videos and interactive tools with the best quality, that students have to master before coming to the class.
Fortunately, the course redesign strategies in flipped classrooms could be synergetic with the recent progress of IT based education. The advent of educational resources of MOOCs like edX, Coursera, etc. not only gives chances for the general public to participate in the higher education but also provides educational resources for university leaders to develop creative flipped learning models \citep{sampson,kate,wulf}.

From its inception in 2009, our university has experimented with the flipped learning model as a possible solution (1) to offer all courses in English, (2) to provide two-way interactive learning for critical thinking and problem solving, and (3) to reduce or at least contain educational costs without sacrificing quality. In line with the university's policy, the introductory physics course has been redesigned for flipped learning as in \citep{bates,chen,stelzer,sadaghiani,sadaghiani2,tomory}.
Our concern is to minimize the lecturing part, which is repeated without substantial change every semester, by replacing it with online self-study before the class and in-class problem solving/discussion.
It may be possible to entirely replace the face-to-face lectures in classroom with the movie clips of famous lecturers, and to complement the class with other effective elements like group problem solving and discussion.

However, the face-to-face lecture and the art of storytelling in class have their own virtues in physics education, especially for the comprehension of physics concepts \cite{redish,jonassen,berger}. First, the instructor can control the pace and amount of the explanations or demonstrations instantly  according to students' responses in the classroom. Second, the instructor can choose the methods and control the time necessary to explain physical phenomena, instantly according to students' responses. In this regard, it is necessary to investigate carefully what will happen to students in flipped classrooms when we reduce the proportion of the face-to-face lecture and replace it with other course elements. Hence, we need to study the impact of the face-to-face lecture on students' learning by systematically adjusting the lecture hours, and this is the main motivation of this work.

The research questions of this article are as follows:
\begin{itemize}
\item Does a flipped classroom offer a more effective way of learning physics than a traditional lecture oriented classroom?
\item Does there exist an optimal proportion of face-to-face lecture compared to interactive in-class activities that replace it?
\item What types of students benefit and suffer most from our course redesign based on flipped learning, due to reduced hours of face-to-face lecture and increased hours of interactive in-class activities such as group problem solving and discussion?
\end{itemize}

\section{Methods}

In this paper, we report achievements and reflections of our students who have taken the first part of introductory university physics - calculus-based classical mechanics and thermal physics, for two semesters - Semester 2012 and 2013. We systematically controlled the proportion of the face-to-face lectures by spending different amounts of lecturing time in different classes, compensating it with other online materials or activities. In the first semester, two instructors joined to teach two different types (traditional and redesigned) of classes separately. In the next semester, one instructor ran two redesigned classes by assigning the proportion of face-to-face lectures differently for the two classes.

\subsection{Semester 2012}

In the first semester, two classes of the first part of the introductory physics course for freshmen were opened.
One instructor decided to apply the redesign method to his class, \textit{Class I} of  $ N_\mathrm{I} = 176$ students,
and the other instructor took traditional method heavily based on lecture in his class, \textit{Class II} of $N_\mathrm{II} = 161$ students.
Since students were assigned to one of the two classes randomly by the university, we could safely assume that initial knowledge of the students about physics was equivalent over those two classes.

For the redesigned class, \textit{Class I}, we decided to take a hybrid model approach, spending half the class time doing face-to-face lectures for summarizing the chapters, and the other half on interactive activities.
Since not all of the contents could be covered by the face-to-face lecture within the limited time,
the core materials for the class, including lecture notes and movie clips, were opened to students on LMS(Learning Management System) before the class.

In class, the instructor gave summary lectures reviewing the main contents of the chapters and emphasizing essential parts.
Often clicker questions were prepared to help the students catch the important concepts \citep{mazur1,mazur2,bruff},
and then group activities were followed.
Students made groups of 2 - 4 members and were allowed to solve the problems selected by the instructor together within the group members.
The instructor showed problems one by one on a big screen
and then gave some time, normally 10 - 15 minutes per problem, for each group to make their own solution.
While students discussed a possible solution to the problem,
the instructor and 2 - 3 TA's walked around the discussion groups, answered students' questions,
and explained some crucial concepts that students might misunderstand.
Then the instructor showed standard solutions to all students while emphasizing important points.

After class, the activities were generically typical.
Homework problems were assigned to students weekly, and the students were required to submit their solutions to TA's.
We arranged recitation classes for students to get help directly from the TA's.
The class format applied to \textit{Class I }consists of mainly three parts as in Table \ref{tb:Iformat}.

In \textit{Class II}, the instructor delivered a traditional type of lecture for the whole class time.
After class, the same homework problems were assigned to students as in \textit{Class I}.
The same format of recitation classes for \textit{Class I} were open to students of \textit{Class II}.

\begin{table*}
\centering
\begin{tabular}{|c|c|}
\hline
Pre-class & lecture notes, movie clips and reading materials \\
\hline
In-class & summary lectures, clicker and group problem solving \\
\hline
Post-class & homework assignments and recitation classes  \\
\hline
\end{tabular}
\caption{The course format of \textit{Class I }, i.e. the flipped classroom, in Semester 2012}{\label{tb:Iformat}}
\end{table*}

We take students' scores of the midterm and final exams as the measure of students' assessment on the learning contents, and compare the score distributions of the two classes by T-test and F-test.

\subsection{Semester 2013}

In the second semester,
we explore the effect of the proportion of the face-to-face lecture on students' learning.
One instructor decided to adopt the same format of course redesign based on the flipped learning to his two classes,
while assigning different proportions of the face-to-face lecture to them.

The course format was similar to that of the redesigned class in Semester 2012, except for two things.
To emphasize students' self-study, we added pre-class quiz assignments based on the contents that students had to prepare before coming to class.
The pre-class quizzes were multiple-choice problems assigned on LMS to be evaluated automatically.
Students had an unlimited number of attempts at solving the problems before class but they were required to achieve at least an 80\% success rate.
Second, though clicker technique is famous as an effective tool to learn physics \citep{mazur1,smith},
we noticed that our students liked to have more time to do group problem solving rather than clicker discussion in Semester 2012.
The reason might be that the instructor and students were not trained enough to make suitable interactions by using clickers.
However, group problem solving was technically easier to be implemented in class and hence we decided not to include clicker questions in Semester 2013.
The class format commonly applied to \textit{Class 1} and \textit{Class 2} consists of mainly three parts as in Table \ref{tb:1and2format}.

\begin{table*}
\centering
\begin{tabular}{|c|c|}
\hline
Pre-class &  lecture notes, movie clips, reading materials and pre-class quiz assignments \\
\hline
In-class &  summary lectures and group problem solving \\
\hline
Post-class & homework assignments and recitation classes  \\
\hline
\end{tabular}
\caption{The class format of \textit{Class 1} and \textit{Class 2} in Semester 2013}{\label{tb:1and2format}}
\end{table*}

The teaching formats of two classes in Semester 2013 are summarized in Table \ref{tb:12}.
The same instructor taught two classes: in \textit{Class 1} of $N_1= 160$ students,
he spent, per week, 1.3 hours for face-to-face lecturing and 2.7 hours for group discussion and problem solving (the proportion of face-to-face lecture is 1/3).
In \textit{Class 2} of $N_2 = 176$ students, he spent, per week, 2 hours for face-to-face lecturing and 2 hours for group discussion and problem solving
(the proportion of face-to-face lecture is 1/2).
There is one important difference between the two classes other than the proportion of  the face-to-face lecture.
In \textit{Class 1}, students were required to come to the first class of the week to participate in the face-to-face lecture and the group activity, but the second class fully dedicated to group problem solving was optional, i.e. students were not required to come to the second class only for group problem solving.
Students had more freedom to choose what would be better for themselves in \textit{Class 1}.
In \textit{Class 2}, however, students were required to come to each of two classes of the week and spent a half of the class time in the face-to-face lecture and the other half in the group problem solving.

\begin{table*}
\centering
\begin{tabular}{|c|c|}
\hline
\hline
choice on the questionnaire & converted points\\
\hline
\hline
strongly agree & +2 \\
\hline
agree & +1 \\
\hline
neutral & 0 \\
\hline
disagree & -1 \\
\hline
strongly disagree & -2 \\
\hline
\hline
\end{tabular}
\caption{Conversion of students' satisfaction of the face-to-face lecture or
group problem solving for Semester 2013.
}{\label{conversion}}
\end{table*}

Right before the final exam, we had a chance to get students' feedback on the educational items adopted in our classes through a questionnaire.
We asked student opinions about the satisfaction with the face-to-face lecture and group problem solving/ discussion with the questions,
``Did the face-to-face lecture contribute much to your learning of physics?'' and
``Did group problem solving contribute much to your learning of physics?''.
The students could choose one among answers, ``strongly agree'', ``agree'', ``neutral'', ``disagree'' or ``strongly disagree''.
The satisfaction level is expressed as an integer number between -2 and 2 which correspond to `strongly disagree' and 'strongly agree', respectively as shown in Table \ref{conversion}.
Though the conversion from the students' answers to the integer numbers between -2 and 2 seem to be arbitrary, we are interested in the statistical features of students' opinions and the absolute value of the conversion is not meaningful to our discussion.

We compare the score distributions of the midterm and final exams of the two classes by T-test and F-test to discuss the impact of changing the proportion of the face-to-face lecture on students' achievements.
We examine the students' degree of satisfaction with face-to-face lecture and group problem solving to explore the role of face-to-face lecturing.
For this, we group students according to midterm and final exam scores, and compare the satisfaction levels of the face-to-face lecture and group problem solving over those groups.
If necessary, we use statistical analysis to judge whether the satisfaction levels of different two groups are meaningfully different.

\section{Results}

\subsection{Results from Semester 2012: Lecture-oriented class vs. Flipped class}

\begin{figure*}
\centering
\includegraphics[scale=0.34]{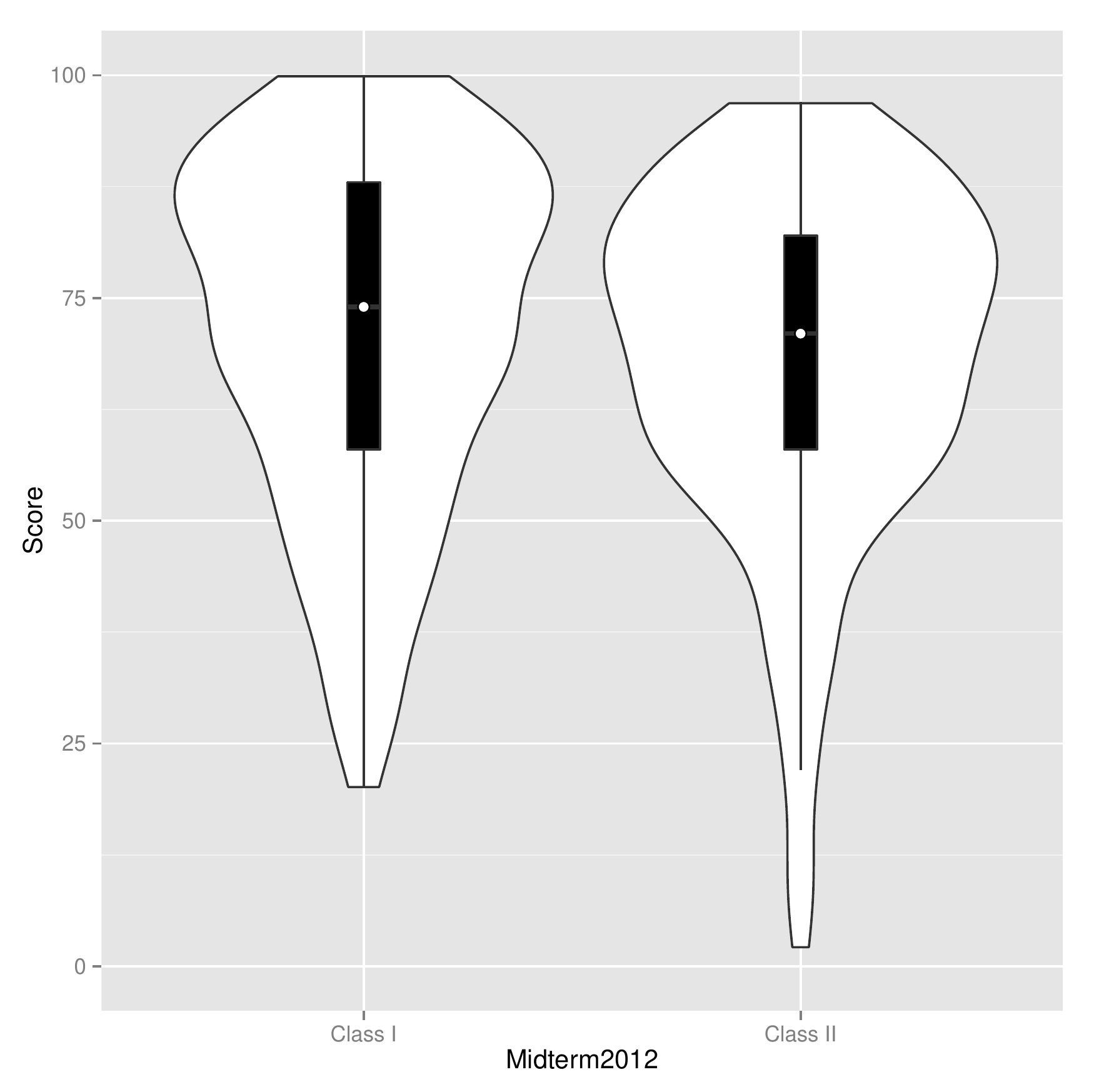}
\includegraphics[scale=0.34]{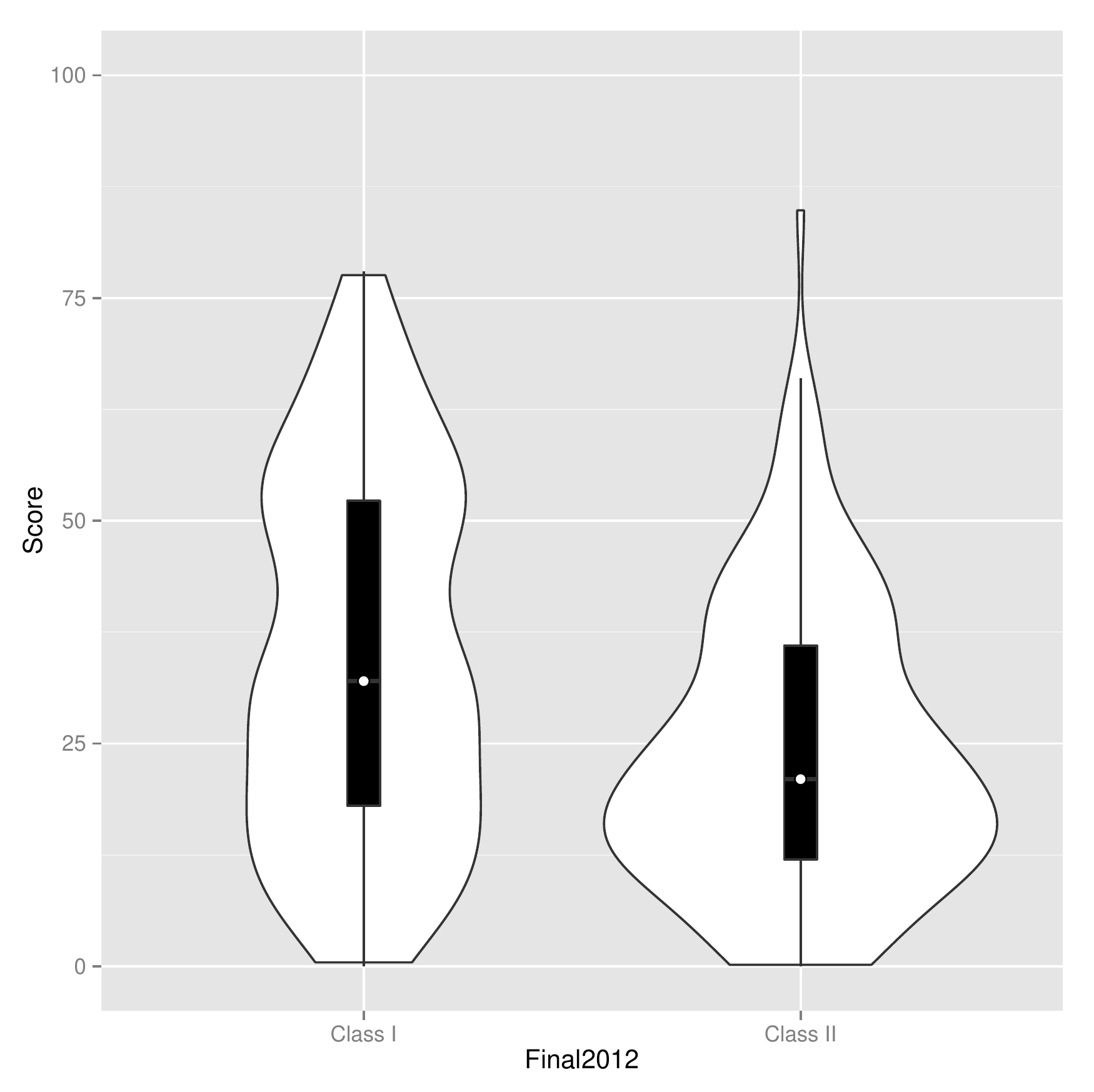}
\caption{ Violin(white region) and box(black box) plots of the scores of \textit{Class I} and \textit{Class II} in the midterm and final exam for Semester 2012.
The widths of the white regions represent the sample frequencies of the students at a given score.
The black boxes represent the groups of the central 50 \% students for the exams, and the white dots within the black boxes are the median scores. }\label{fig:IandII}
\end{figure*}

\begin{table*}
\centering
\begin{tabular}{|c|c|c|}
\hline
\hline
&\textit{Class I}(redesigned) & \textit{Class II} (traditional) \\
\hline
\hline
number of students & $ N_\mathrm{I} = 176$ & $N_\mathrm{II} = 161$ \\
\hline
midterm exam & 71.4 ($\pm$ 1.48)    &  68.2 ($\pm$ 1.51)   \\
\hline
final exam &  34.4 ($\pm$ 1.54)      &  23.9 ($\pm$ 1.88)  \\
\hline
\hline
\end{tabular}
\caption{Number of students and average exam score ($\pm$ its standard error, i.e. the standard deviations divided by the square root of the sample numbers ) in classes for Semester 2012}{\label{tb:IandII}}
\end{table*}

\begin{table*}
\centering
\begin{tabular}{|c|c|c|c|c|}
\hline
\hline
& T-test & Degree of freedom & p-value & 95\% confidence level  \\
\hline
\hline
Midterm exam & 1.5146 & 335 & 0.1308 & -0.9578, 7.3703  \\
\hline
Final exam & 5.3203 & 327.315 & 1.925 $\times 10^{-7}$ & 6.663906, 14.483327 \\
\hline
\hline
& F-test & Degree of freedom & p-value & 95\% confidence level \\
\hline
\hline
Midterm exam & 1.0374 & 175/160 & 0.8144 & 0.7645, 1.4048 \\
\hline
Final exam & 1.6324 & 175/160 & 0.0017 & 1.2030, 2.2105 \\
\hline
\hline
\end{tabular}
\caption{T-test and F-test for the comparison of the exam scores for
\textit{Class I} and \textit{Class II} in Semester 2012.}{\label{tb:TF1}}
\end{table*}

Though the course formats of \textit{Class I} and \textit{Class II} were different from each other,
we gave exactly the same problems in the midterm and final exams, to compare the effectiveness of students' learning in both classes.
The exam problems were consisted of multiple choice or simple answer questions (30\%) and  numerical questions (70\%).
We included challenging questions to control the overall difficulty of the exams.

The midterm exam was about Newton mechanics, the contents covered in the first 6 weeks.
It is found that the average scores of the midterm exam for both classes are similar as in Table \ref{tb:IandII}.
The final exam was about rotational motion, wave dynamics and thermal physics, the contents covered in the last 6 weeks.
Since the average score of the midterm exam was higher than we desired, we decided to include more challenging questions in the final exam.
The average scores of the final exam for both classes seemed to decrease substantially in comparison with the midterm scores in Table \ref{tb:IandII}, because the topics covered after the midterm exam were new and are challenging to the students.
The average score of \textit{Class I} was 10 \% higher than that of \textit{Class II}.

The students' populations of the both classes over the score of exams are presented by
the violin plots of the students and the box plots of the central 50\% students in Fig.  \ref{fig:IandII}.
The population within the central 50\% for the midterm score of  \textit{Class I} is slightly more broadly distributed than that of \textit{Class II}.
The similar pattern is to be seen more explicitly in the final exam.
The central 50\% students of \textit{Class I} did get explicitly higher scores than those of \textit{Class II} as in the right panel of Fig. \ref{fig:IandII}.

The statistical properties of \textit{Class I} and \textit{Class II} are checked by T-test and F-test as summarized in  Table \ref{tb:TF1}.
For the midterm exam, the hypothesis that the means of the two classes are equal is accepted by T-test whose p-value is 0.1308 at 5\% of the significant level for two-sided test.
Here we assume that the variances do not differ too much,
as it is confirmed by F-test whose p-value is 0.8144.
Hence, we do not see a significant difference in the means of the two classes for the midterm exam.
However, for the final exam, different results could be noticed.
We use Welch two sample T-test because the variances of two classes are not equal,
as it is approved by F-test whose p-value is 0.0017 at 5\% of the significance level.
The difference in the means of two classes for the final exam is confirmed by T-test, since the p-value is $1.925 \times 10^{-7}$ which is much less than the significant level.

The above analysis indicates that there was no essential difference in students' learning by the midterm exam in both classes but were remarkable changes in the students' performance in the final exam.
One may interpret this result in the following two ways.
First, the freshmen were familiar with the topics covered in the first part before the midterm exam because they already learned most of the topics in their high schools.
Hence, we could conclude that our class redesign seems to be effective in students' learning,
especially when students started dealing with relatively new topics.
In the redesigned class, many of students seemed to have chances to correct their knowledge and understanding on the new topics, while they have time to interact with their peers.
Second, it takes some amount of time for students to get familiar with the new teaching method. Students were usually taught in lecture-oriented classrooms before entering the university and hence needed to be exposed to the redesigned course format to be effective in their learning.
In that sense, it seems natural to see that the effect of course redesign was manifested after the midterm exam.

\subsection{Results from Semester 2013}

\subsubsection{Exam results}

\begin{table*}
\centering
\begin{tabular}{|c|c|c|}
\hline
\hline
& Class 1 & Class 2 \\
\hline
\hline
number of students &  $ N_1$ = 160 & $N_2 $= 176  \\
\hline
time for face-to-face lecture per week & 1.3 hours  & 2.00 hours \\
\hline
time for group problem solving per week & 2.7 hours  & 2.00 hours \\
\hline
portion of face-to-face lecturing & 1/3 & 1/2  \\
\hline
midterm exam & 55.9 ($\pm$ 18.7)   &  56.7 ($\pm$ 17.7)    \\
\hline
final exam &  43.6 ($\pm$ 21.1)    &  47.4 ($\pm$ 21.4)   \\
\hline
\hline
\end{tabular}
\caption{Number of students and average exam score ($\pm$ its standard error, i.e. the standard deviations divided by the square root of the sample numbers) in classes for Semester 2013}{\label{tb:12}}
\end{table*}

\begin{table*}
\centering
\begin{tabular}{|c|c|c|c|c|}
\hline
\hline
& T-test & Degree of freedom & p-value & 95\% confidence level \\
\hline
\hline
Midterm exam & -0.42 & 334 & 0.6747 & -4.740380, 3.072198  \\
\hline
Final exam & -1.6629 & 334 & 0.09728 & -8.4316737, 0.7066737  \\
\hline
\hline
& F-test & Degree of freedom & p-value & 95\% confidence level \\
\hline
\hline
Midterm exam & 1.1061 & 159/175 & 0.514 & 0.8164597, 1.5018120 \\
\hline
Final exam 1 & 0.9757 & 159/175 & 0.8758 & 0.7201962, 1.3247431 \\
\hline
\hline
\end{tabular}
\caption{T-test and F-test for the comparison of exam scores for
\textit{Class 1} and \textit{Class 2} in Semester 2013.}{\label{tb:anova2013}}
\end{table*}

\begin{figure*}[ht]
\centering
\includegraphics[scale=0.34]{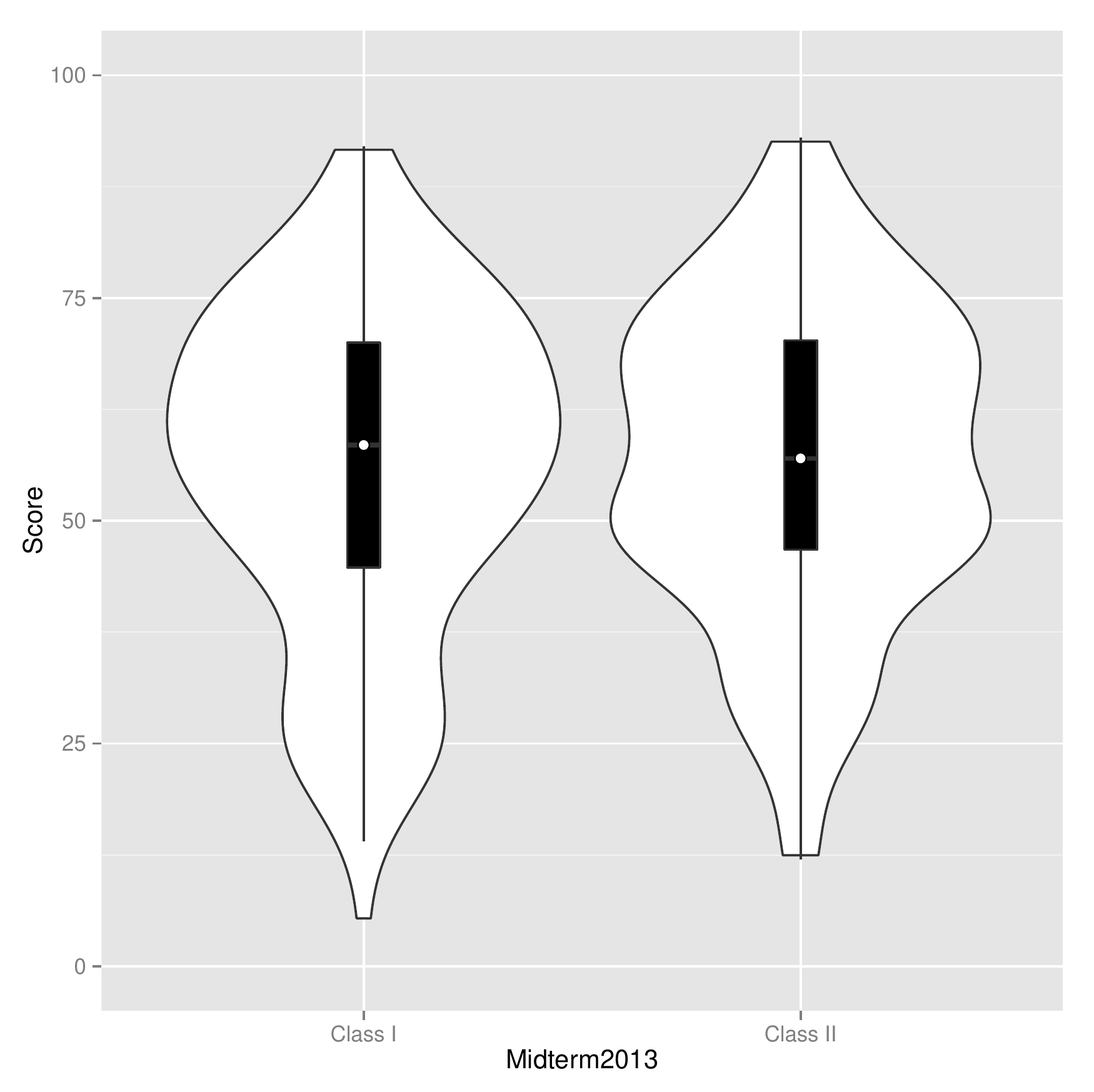}
\includegraphics[scale=0.34]{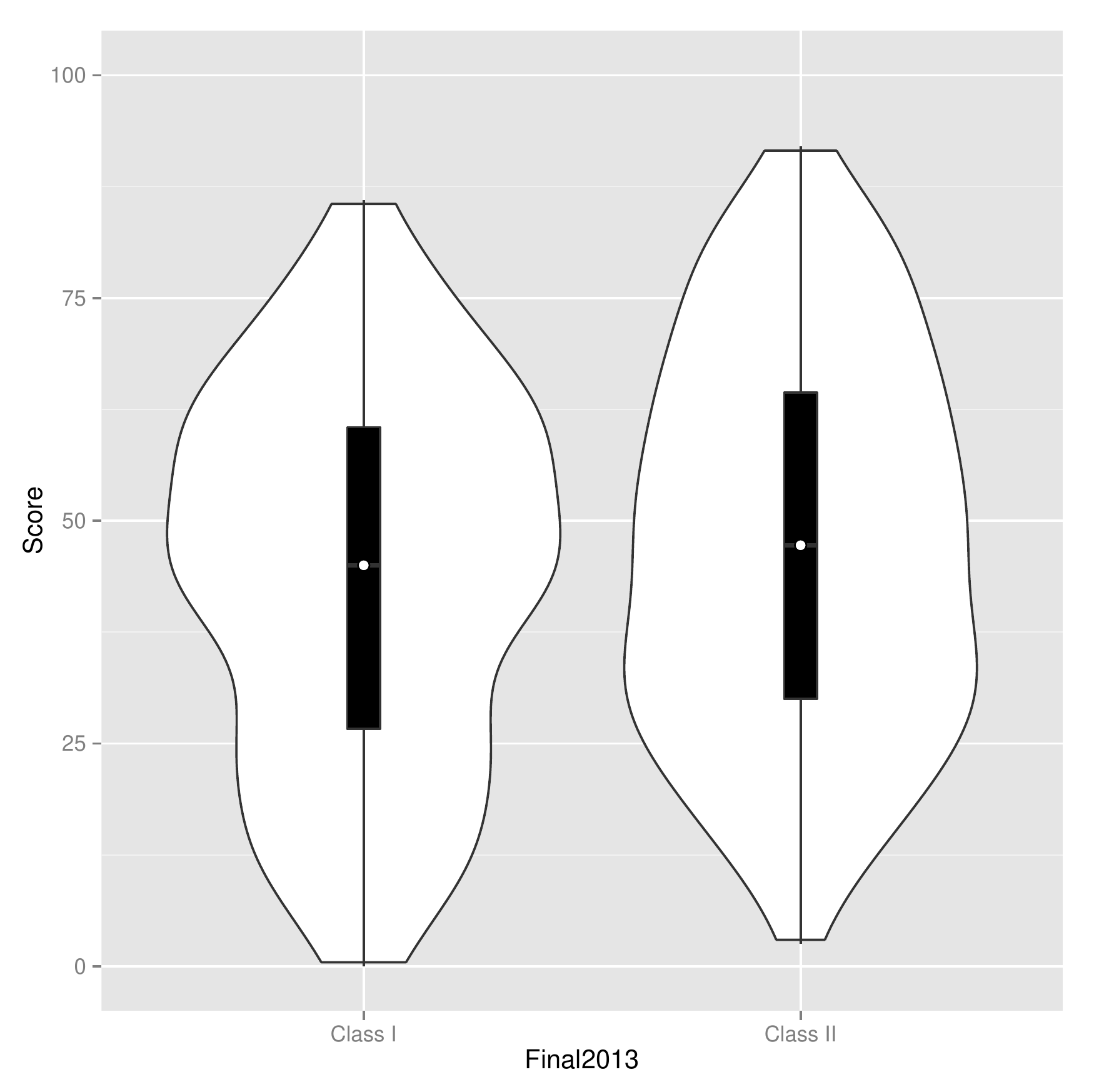}
\caption{ Violin(white region) and box(black box) plots of the scores of \textit{Class 1 } and \textit{Class 2} in the midterm and final exam for Semester 2013.
The widths of the white regions represent the sample frequencies of the students at a given score.
The black boxes represent the groups of the central 50 \% students for the exams, and the white dots within the black boxes are the median scores.
 }\label{fig:12}
\end{figure*}

The exam results of the two classes are summarized in Table \ref{tb:12}, and their violin and box plots are shown in Fig. \ref{fig:12}.
For the midterm exam,
the average scores of \textit{Class 1} and \textit{Class 2} are 55.9 and 56.7 respectively.
For the final exam,
the average scores of the final exam for \textit{Class 1} and \textit{Class 2} are 43.6 and 47.4 respectively.
It seems that students of \textit{Class 2} did perform slightly better than those of \textit{Class 1} in the final exam.
In the violin and box plots of Fig. \ref{fig:12}, the central 50 \% students of \textit{Class 2} got a little higher scores than those of \textit{Class 1}.

However, the results of T-test and F-test imply that the statistical features for the two classes are not different in the midterm and final exam as in Table \ref{tb:anova2013}.
It is noticed that the variances of the two classes for the midterm and final exams do not differ by F-test whose p-values are 0.514 for the midterm exam and 0.8758 for the final exam.
According to T-test under the equal variance assumption,
the p-values are 0.6747 for the midterm exam and 0.09728 for the final exam.
Hence, it is hard to say that the overall statistical distributions for the midterm and final exams are different.

In the setting of our course redesign, the proportion of face-to-face lecture dose not give a profound impact on the students' learning process.
It seems that the lecturing part has a function of specifying the items that students should learn, but the learning outcome is directly related with students' assimilation on those items.
While problem solving and group discussion would be an effective tools for the students' assimilation, the proportions of 2/3 (\textit{Class 1}) and 1/2 (\textit{Class 2}) for such activities seem to be operating equally well with our course redesign.

\subsubsection{Students' reflections on L and GPS}

Now we discuss the students' satisfaction levels with the face-to-face lecture (L) and group problem solving (GPS) from the questionnaire taken right before the final exam.
We asked ``Did the face-to-face lecture contribute much to your learning of physics?'' and ``Did group problem solving contribute much to your learning of physics?'' to the students through the questionnaire.
For the face-to-face lecture, the percentages of students who voted to
`strongly agree' or `agree' are
$77.3\%$ ($=27.3\% + 50.0\%$ ) for \textit{Class 1} and $72.6\%$ ($=22.6\% + 50.0\%$) for \textit{Class 2}, respectively.
For the group problem solving, the percentages of students who voted to
`strongly agree' or `agree' are  $79.7\%$ ($=36.7\% + 43.0\%$) for \textit{Class 1} and $80.5\%$ ($=31.1\% + 49.4\%$ ) for \textit{Class 2}, respectively.
The average satisfaction scores of L and GPS and their standard errors are shown in Table \ref{tb:12LGPS}.
It seems that the students in the both classes generally prefer GPS as much as L.

As summarized in Table \ref{tb:LGPS}, the average satisfaction levels of L or GPS seem to be similar by T-test and F-test.
For L, the variance of \textit{Class 1} is different from that of \textit{Class 2} since the p-value is 0.03056 at the significance level of 5\%.
Then we notice that the p-value of Welch two sample T-test is 0.8256 which is bigger than the significance level.
For GPS, the variances of the two classes seem to be similar since the p-value of F-test and T-test are 0.838 and 0.4911 respectively.

\begin{table*}
\centering
\begin{tabular}{|c|c|c|}
\hline
\hline
& \textit{Class 1} & \textit{Class 2} \\
\hline
\hline
face-to-face lecture  &  $0.937$ ($\pm 0.078$)     &    $ 0.951$ ($\pm 0.058$)   \\
\hline
group problem solving &  $1.118$ ($\pm  0.076$   )   &  $1.067$ ($\pm 0.065$)  \\
\hline
\hline
\end{tabular}
\caption{Average satisfaction levels converted into scores by Table \ref{conversion} ($\pm$ its standard error, i.e. the standard deviations divided by the square root of the sample numbers) in classes for Semester 2013}{\label{tb:12LGPS}}
\end{table*}

\begin{table*}
\centering
\begin{tabular}{|c|c|c|c|c|}
\hline
\hline
& T-test & Degree of freedom & p-value & 95\% confidence level \\
\hline
\hline
L & 0.2205 & 284 & 0.8256 & -0.1688569, 0.2114694 \\
\hline
GPS &-0.6895 &  284  & 0.4911
 & -0.2667435, 0.1283524\\
\hline
\hline
& F-test & Degree of freedom & p-value & 95\% confidence level\\
\hline
\hline
L & 0.6951 & 123/161 & 0.03056 & 0.4962111, 0.9664543\\
\hline
GPS & 0.9672 & 123/161 & 0.838 & 0.6904185, 1.3447059\\
\hline
\hline
\end{tabular}
\caption{T-test and F-test for
the comparison between the students' satisfaction with L and GPS of \textit{Class 1} and \textit{Class 2} in Semester 2013.
}{\label{tb:LGPS}}
\end{table*}

In the next sections, we look into the satisfaction levels of L and GPS by grouping students' scores of the midterm exam, the final exam and combination of both.
The objective is to find student groups to which instructors should pay careful attention when they optimize the proportion of the face-to-face lecture in flipped learning.

\subsubsection{For students who performed well in the midterm and final exams}

We propose, as a measure of the students' performance, the algebraic sum,
\begin{equation}
\centering
\mathrm{M+F} \equiv \mathrm{midterm~exam~score} + \mathrm{final~exam~score}~.
\end{equation}
M+F is the total score that each student gained from the midterm and final exams, regardless of how much the final exam score was improved or not, compared to the midterm score for a student.

\begin{figure*}[ht]
\centering
\includegraphics[scale =0.37]{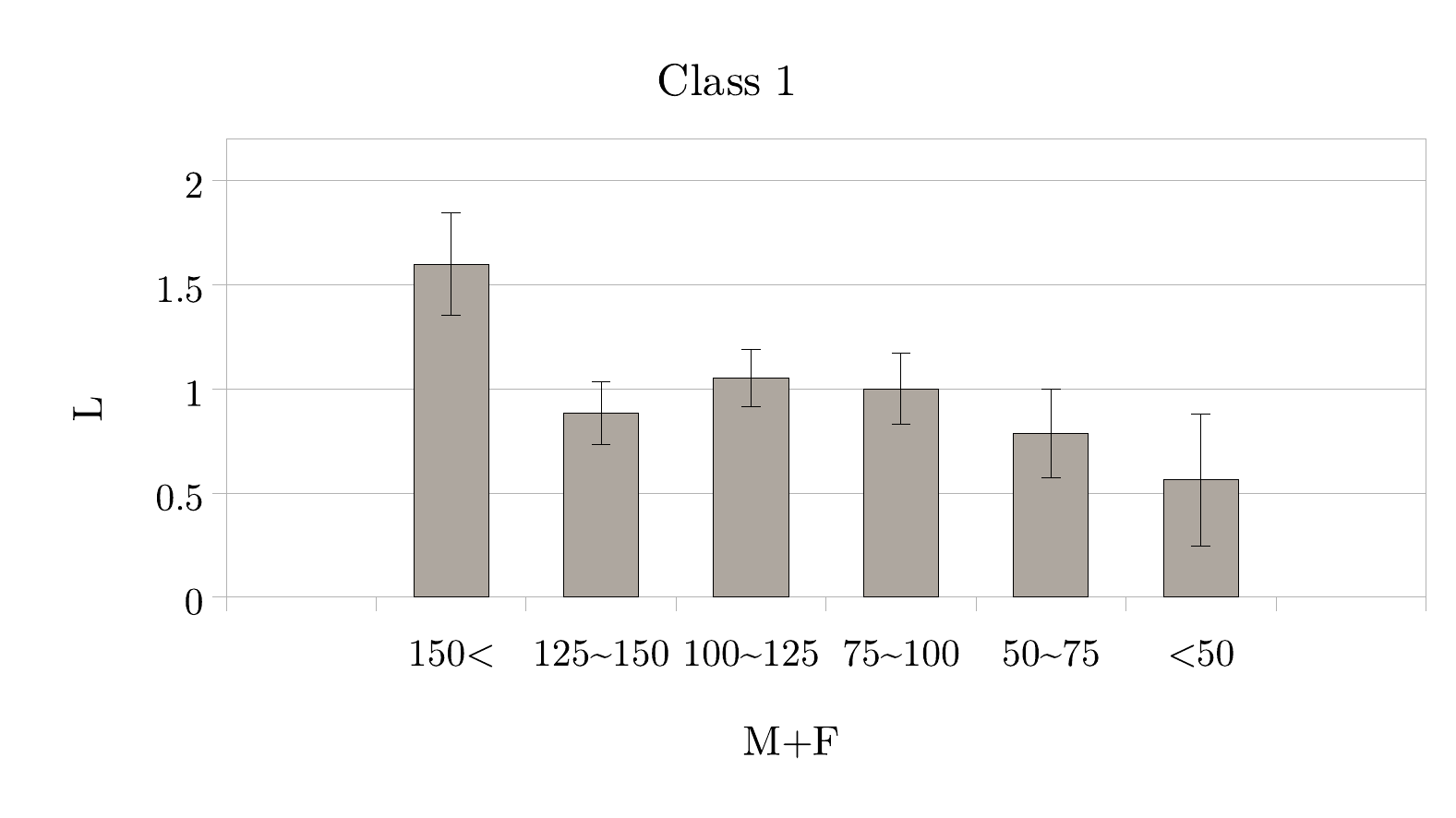}
\includegraphics[scale =0.37]{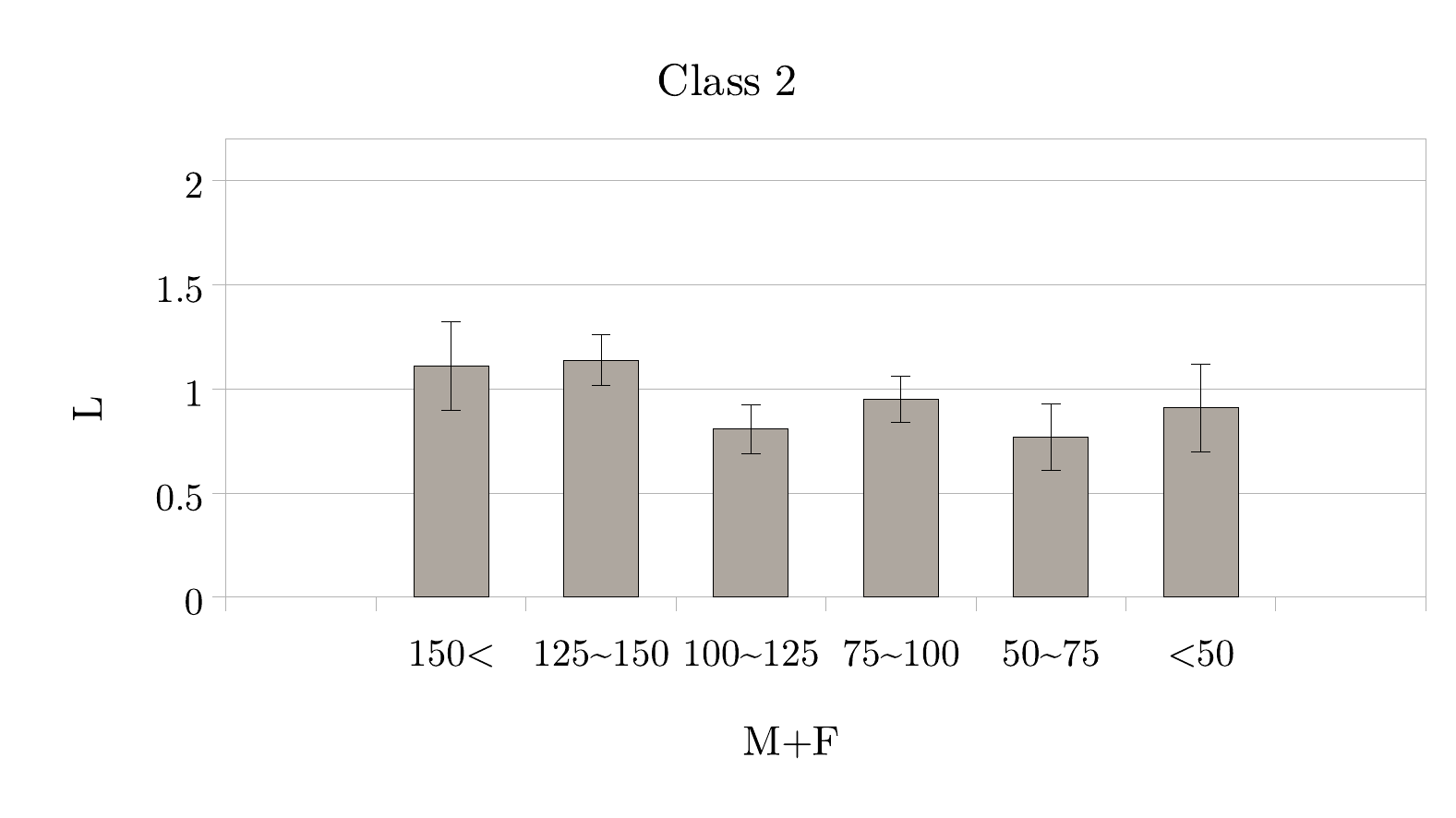}
\includegraphics[scale =0.37]{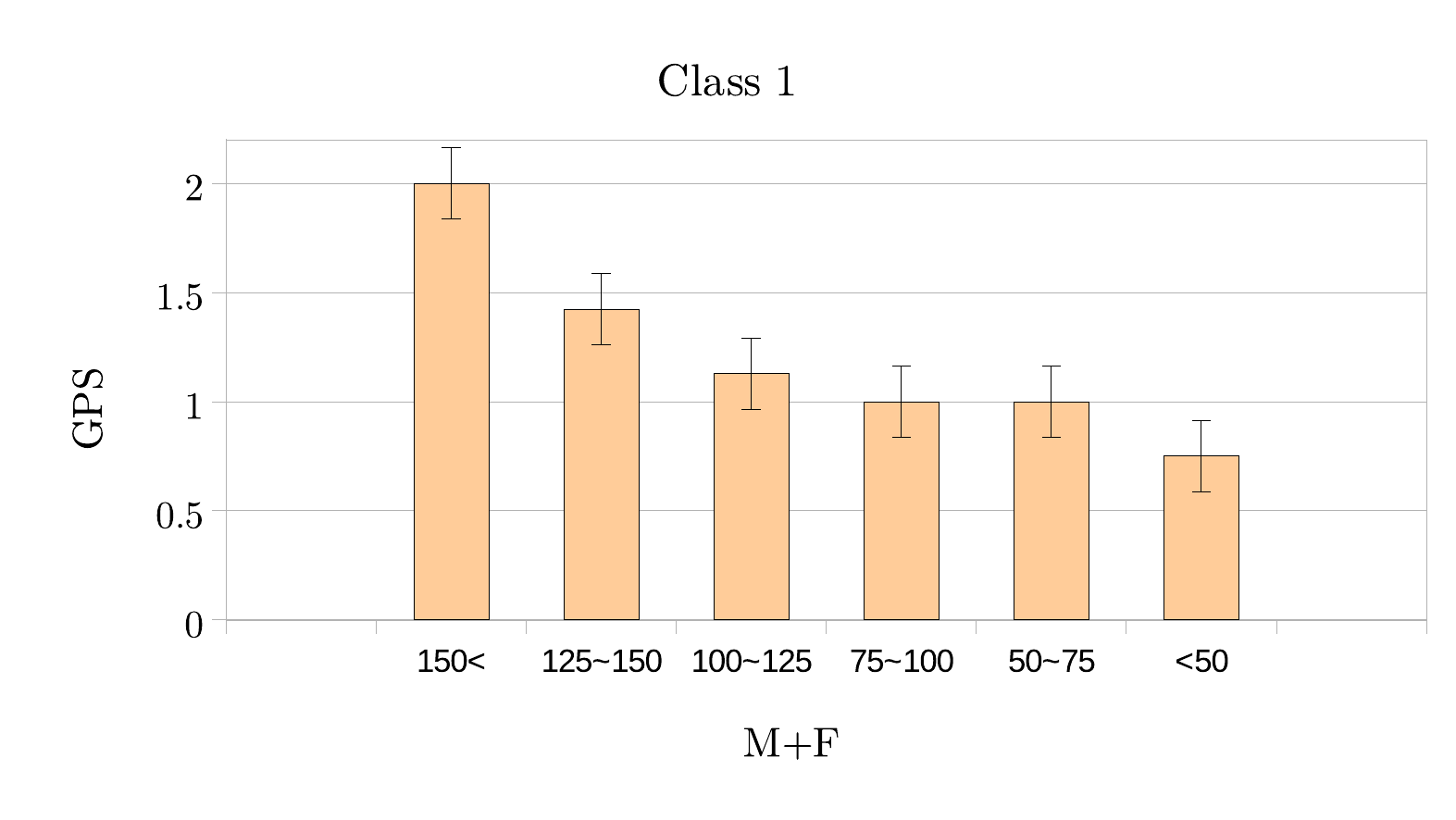}
\includegraphics[scale =0.37]{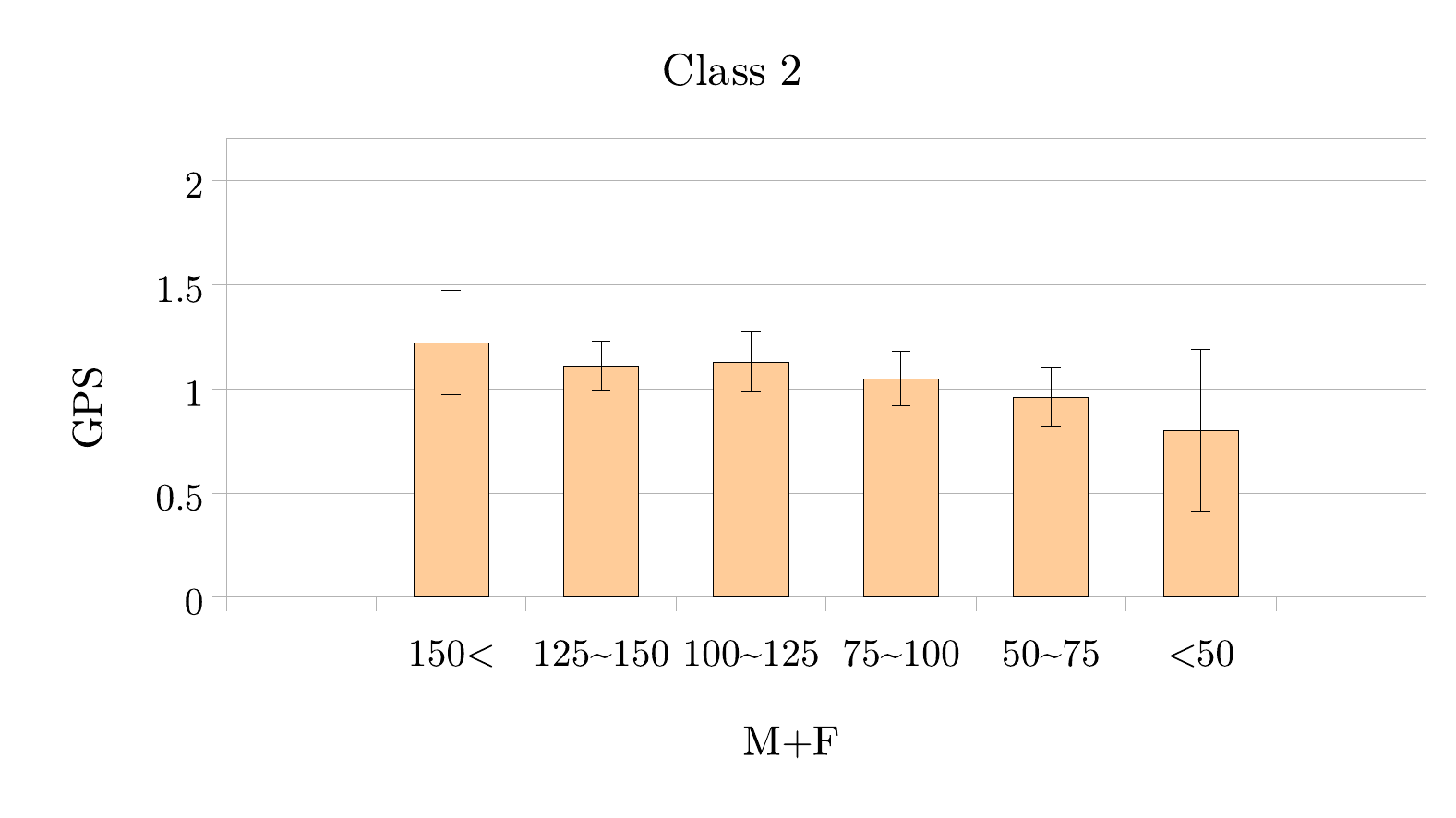}
\caption{The students' satisfaction levels with lecture(L) or to group problem solving (GPS) versus the ranges of the sum of midterm and final exam score(M+F) for \textit{Class 1} and \textit{Class 2} in Semester 2013.
}\label{fig:m+f}
\end{figure*}

The students in each of \textit{Class 1} and \textit{Class 2} are grouped according to their M+F scores into `below 50', `50 to 75', `75 to 100', `100 to 125', `125 to 150' and `higher than 150',
and then the satisfaction level of each group with the face-to-face lecture(L) and group discussion and problem solving (GPS) is analysed to compare \textit{Class 1} and \textit{Class 2}.
Fig. \ref{fig:m+f} shows the average satisfaction scores, and the standard errors, i.e. the standard deviations divided by the square root of the sample numbers for each M+F group.

\begin{table*}
\centering
\begin{tabular}{|c|c|c|c|c|}
\hline
\hline
& T-test & Degree of freedom & p-value & 95\% confidence level \\
\hline
\hline
L & -0.6157 & 85 & 0.5397 & -0.4531226, 0.2388369 \\
\hline
GPS & 2.0437 & 85 & 0.04408 & 0.008642709, 0.701186784\\
\hline
\hline
& F-test & Degree of freedom & p-value & 95\% confidence level\\
\hline
\hline
L & 0.6709 & 30/55 & 0.2391 & 0.3649576, 1.3101065\\
\hline
GPS & 0.9893 & 30/55 & 0.9986 & 0.5381823, 1.9319399\\
\hline
\hline
\end{tabular}
\caption{T-test and F-test for
the comparison between the student groups with $ \operatorname{M+F} >125$ from \textit{Class 1} and \textit{Class 2} in Semester 2013.
}{\label{tb:anova2013mf}}
\end{table*}

The upper two panels of Fig. \ref{fig:m+f} show that the satisfaction levels of the face-to-face lecture (L) for the two classes are not different within the standard errors, except for the students whose M+F is higher than 150 in \textit{Class 1}, while the sample size is too small, 5.
Hence, one can say that the preference to the face-to-face lecture is similar over the whole M+F groups.
However, the lower two panels of Fig. \ref{fig:m+f} show that the satisfaction levels of group problem solving (GPS) for the students with $\operatorname{M+F} > 125$ of \textit{Class 1} is higher than that of \textit{Class 2}, while the sample sizes are 31 for \textit{Class 1} and 54 for \textit{Class 2}, respectively.

We need to carefully check the statistical properties of the student groups of $\operatorname{M+F} > 125$ from \textit{Class 1} and \textit{Class 2} by doing T-test and F-test.
In Table \ref{tb:anova2013mf}, the variances of the satisfaction levels of L and GPS seem to be equal as addressed by F-test whose p-values are 0.2391 for L and 0.9986 for GPS at 5\% of the significance level.
The results of T-test of the satisfaction levels of L and GPS for the two groups are significantly different.
For L, it is noticed that the means of the satisfaction level do not differ by T-test whose p-value is 0.5397.
However, for GPS, the p-value of the T-test is 0.04408 which implies that the two groups have different preferences to the group activity.
It means that while the statistical distributions of the samples for the face-to-face lecture are not different, those for the group problem solving are distinguishable.
Hence, one may conclude that students with $\operatorname{M+F} > 125$ of \textit{Class 1} tended to think that group problem solving was helpful in their learning of physics rather than those of \textit{Class 2}.

It seems that the comparison between the students with $\operatorname{M+F} > 125$ of \textit{Class 1} and \textit{Class 2} shows that GPS can be a relevant substitute of L for the students who performed well.
Since the instructor had limited time to cover the contents in \textit{Class 1} rather than in \textit{Class 2}, the \textit{Class 1} is a test ground where the students notice whether group problem solving could properly compensate the reduction of face-to-face lecturing.
For those students of \textit{Class 1} who could easily follow up the pace of the course, it seemed clear that interactive engagements like group problem solving was carefully designed to convey all the contents and to help students assimilate them.

\subsubsection{For students whose exam scores were reduced }

We could conclude that the overall students' performance of both classes were similar from Table \ref{tb:anova2013}.
However, when we compare the positions of the central 50 \% groups for the midterm and the final exams in Fig. \ref{fig:12},
the overall change in the scores of the central 50 \% group of \textit{Class 1} seems to be larger  than that of \textit{Class 2}.
It may be meaningful to trace students' improvements by comparing their scores of the midterm exam with those of the final exam to see the relations with the face-to-face lecture and group problem solving.

Since the distribution with the large number of samples like the student numbers of our classes would be Gaussian,
the score of the final exam for each student can be renormalized
so that the average and standard deviation of the final exam are same as those of the midterm exam in the class where the student belongs.
The renormalized final exam score is
\begin{equation}
\centering
\operatorname{FR} \equiv  \left( \mathrm{F} - \mathrm{FA}   \right)\frac{\sigma_{M}}{\sigma_{F}} + \mathrm{MA}
\end{equation}
where $\mathrm{MA}$, $\mathrm{FA}$, $\sigma_{M}$ and $\sigma_{F}$ are the average scores and standard deviations of the midterm and final exams for each of \textit{Class 1} and \textit{Class 2}.
Then, the improvement of a student in a class could be quantified by comparing the midterm exam score and the renormalized final exam score.
Now, the improvement factor for each student in one class, FR-M is introduced by
\begin{equation}
\operatorname{FR-M} \equiv \operatorname{FR}-\operatorname{M}= \left[ \left( \mathrm{F} - \mathrm{FA}   \right)\frac{\sigma_{M}}{\sigma_{F}} + \mathrm{MA}  \right] - \mathrm{M}
\end{equation}\label{eq:frm}
where M is the midterm exam score of the student.

The students in each of \textit{Class 1} and \textit{Class 2} are grouped by their FR-M scores
into `below -10', `-10 to 0', `0 to 10', and `higher than 10',
and then the average and standard error of the satisfaction level with L or GPS for each FR-M group is analysed as it is done for M+F.

\begin{figure*}[ht]
\centering
\includegraphics[scale =0.37]{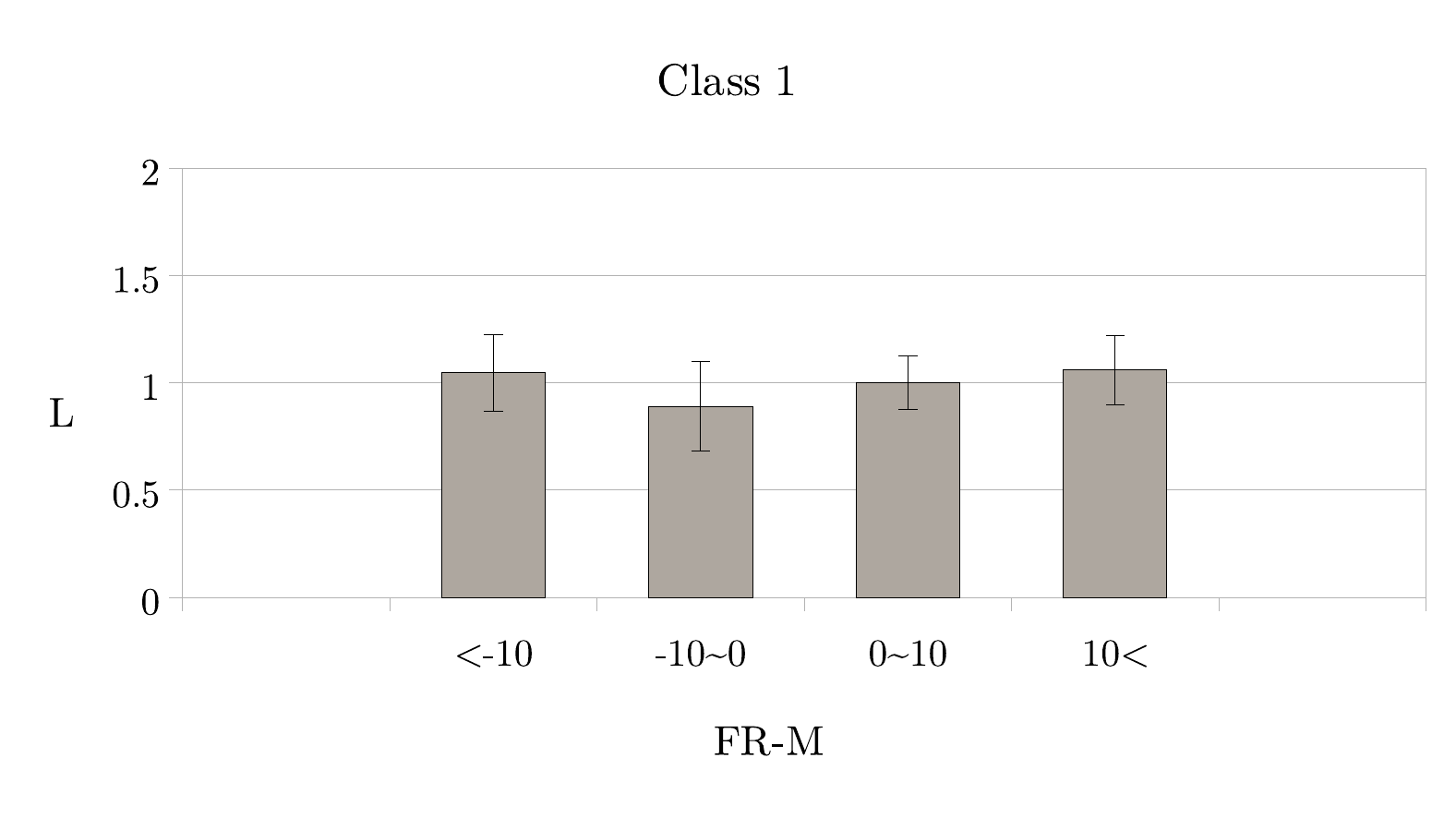}
\includegraphics[scale =0.37]{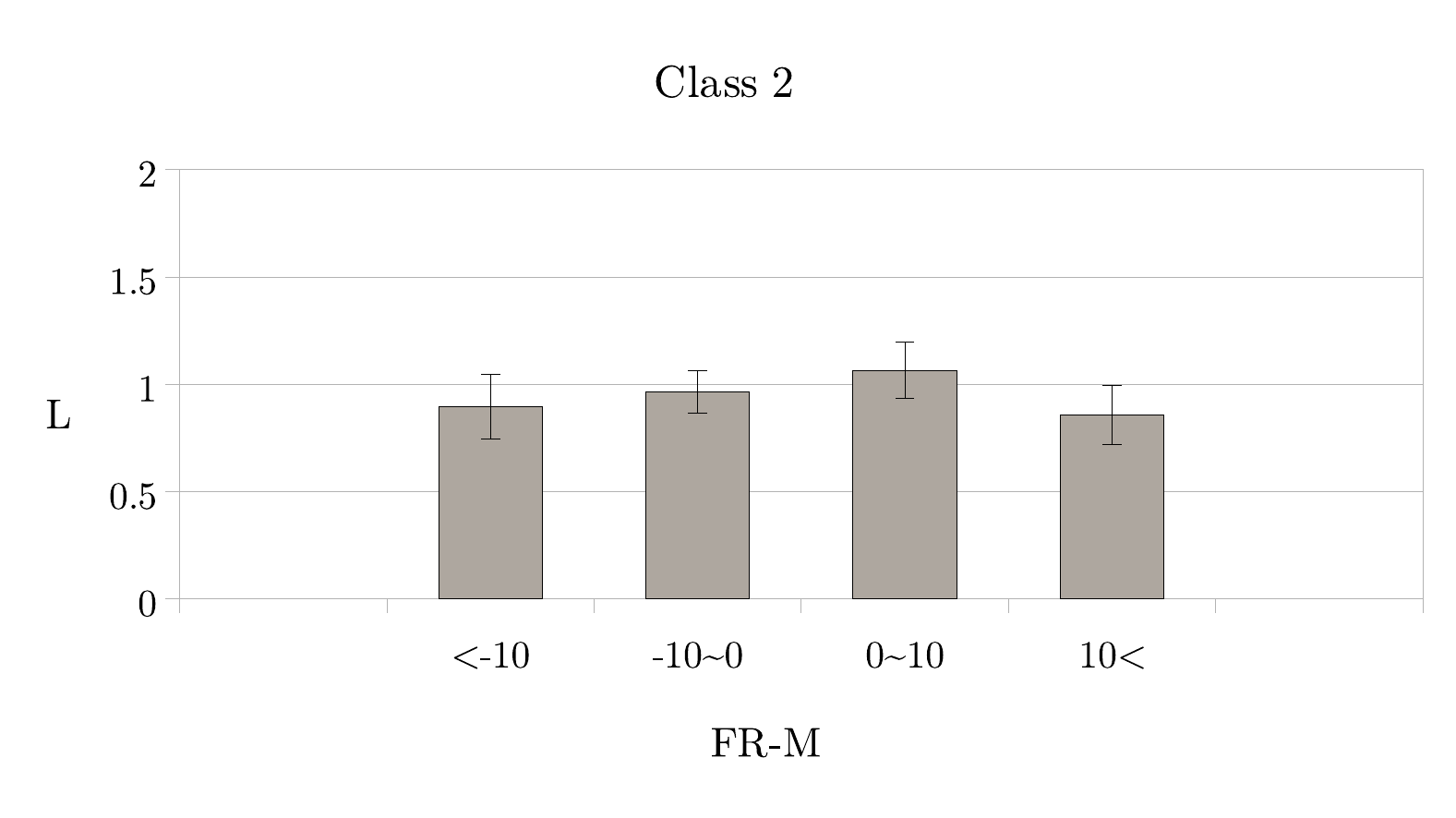}
\includegraphics[scale =0.37]{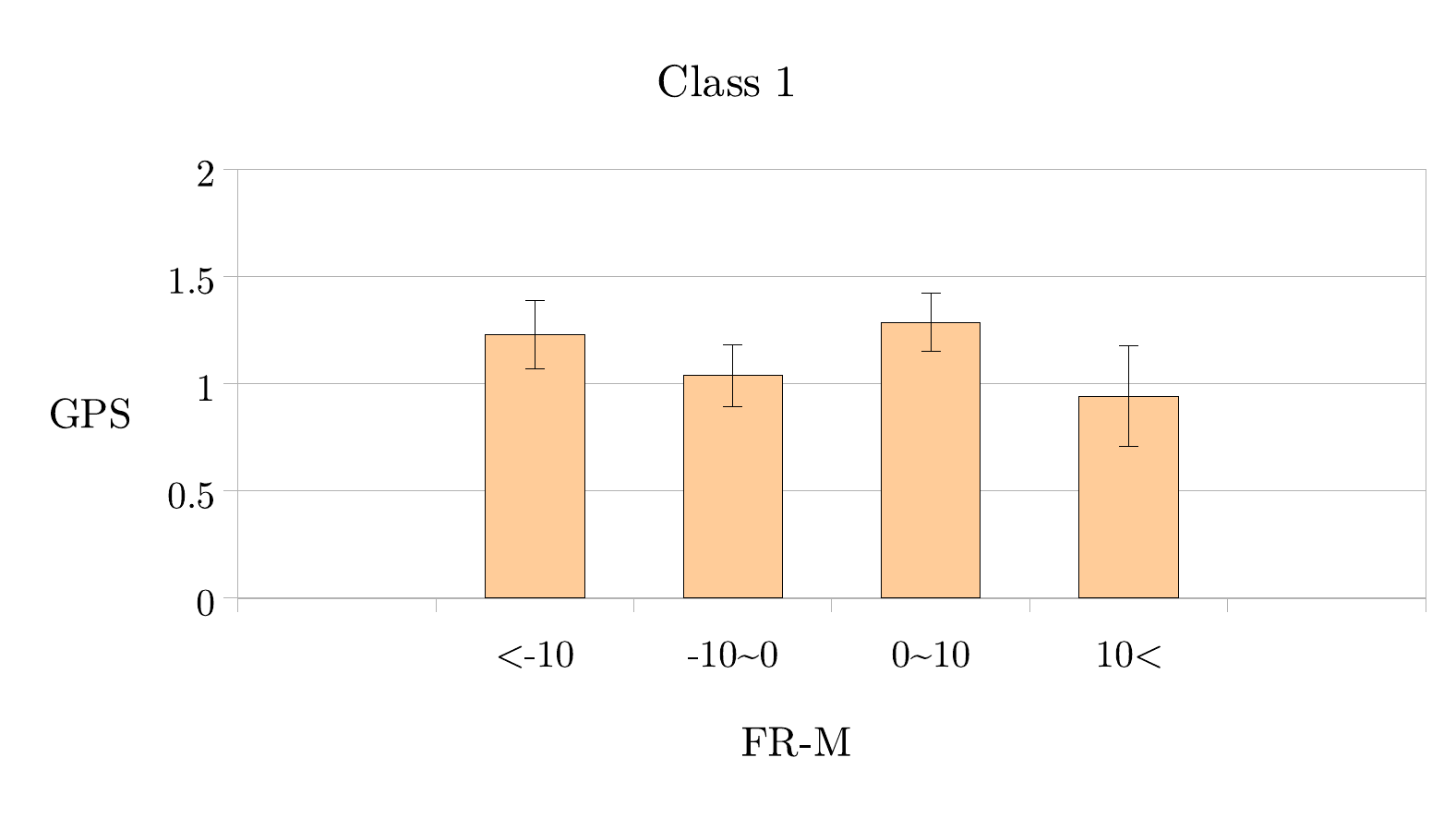}
\includegraphics[scale =0.37]{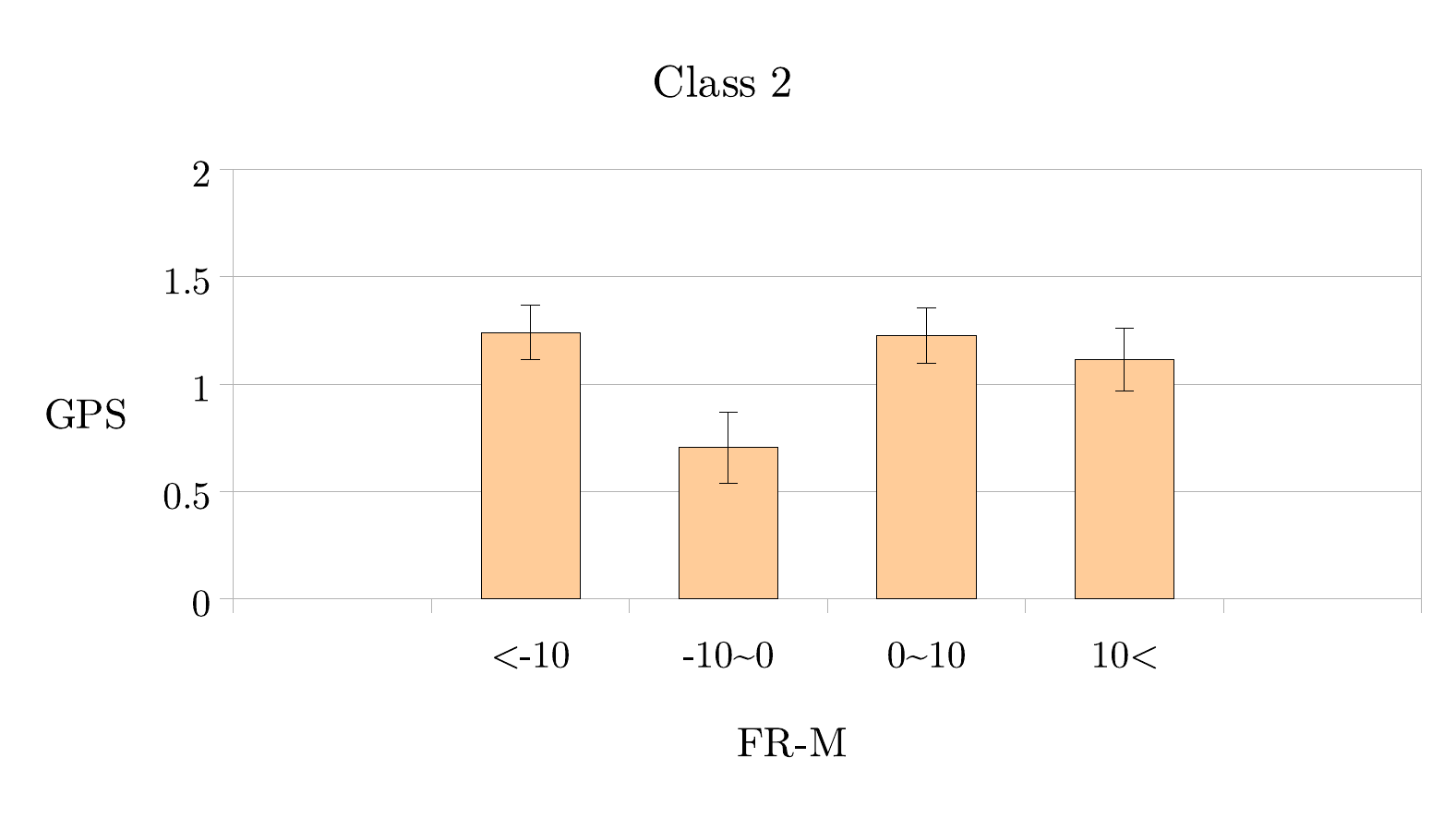}
\caption{The students' satisfaction levels with lecture(L) or to group problem solving(GPS) versus the ranges of the relative improvement in the score of the final exam to that of the midterm exam(FR-M) for \textit{Class 1} and \textit{Class 2} in Semester 2013.
}\label{fig:fr-m}
\end{figure*}

\begin{table*}
\centering
\begin{tabular}{|c|c|c|c|c|c|}
\hline
\hline
& T-test & Degree of freedom & p-value & 95\% confidence level\\
\hline
\hline
L & -0.8091 & 49.804 & 0.4223 & -0.6177194, 0.2629957\\
\hline
GPS & 1.9072 & 65 & 0.06091 & -0.02126542, 0.92322621\\
\hline
\hline
& F-test & Degree of freedom & p-value & 95\% confidence level\\
\hline
\hline
L & 3.7361 & 33/32 & 0.0003272 & 1.851972, 7.508512\\
\hline
GPS & 0.9139 & 33/32 & 0.7975 & 0.4530013, 1.8366181\\
\hline
\hline
\end{tabular}
\caption{T-test and F-test for the comparison between the student groups with $ -10 < \operatorname{FR-M} < 0$ from \textit{Class 1} and \textit{Class 2} in Semester 2013.}{\label{tb:anova2013frm}}
\end{table*}

The upper two panels of Fig. \ref{fig:fr-m} show the satisfaction levels with L for the two classes are not different over the whole FR-M range.
However, the lower two panels of Fig. \ref{fig:fr-m} show
the satisfaction level of the students with $ -10 < \operatorname{FR-M} < 0$ in \textit{Class 2}
seems to be relatively low compared to other FR-M ranges.
The students with $ -10 < \operatorname{FR-M} < 0$ in \textit{Class 2}  tend to think that they do not get helps from the group problem solving as much as from the face-to-face lecture, rather than in \textit{Class 1}.

Now we take the samples of $ -10 < \operatorname{FR-M} < 0$ separately from two classes to do T-test and F-test. The sample sizes are commonly 34 for the two classes.
In Table \ref{tb:anova2013frm},
while the variances of L seem to be different since the p-value of F-test is 0.0003272,
those of GPS do not differ since the p-value is 0.7975.
For the face-to-face lecture, Welch Two sample test is used and its p-value is 0.4223 which implies the mean values of L for the two groups are likely to be same.
However, for the group problem solving, T-test under the equal variance assumption is used and its p-value is 0.06091.
If we consider a less conservative significant level, i.e. at 10\% or higher,
then this test for GPS is rejected clearly.
It means that the students in \textit{Class 2}
whose final exam scores are slightly reduced compared to the midterm exam scores preferred not the group problem solving, rather than in \textit{Class 1}.

In the groups with $ -10 < \operatorname{FR-M} < 0$,
there are 43 students in \textit{Class 1} and 36 in \textit{Class 2}.
The average score of those students are reduced from 59.4 ($\pm$ 2.64, the standard error)  to 43.2 ($\pm$ 3.12, the standard error) for \textit{Class 1} and from 58.6 ($\pm$ 2.53, the standard error) to 42.3 ($\pm$ 3.02, the standard error) for \textit{Class 2}.
The students in these groups belong to the central 50 \% groups in the midterm and final exams, which motivate us to introduce the parameter $\operatorname{FR-M}$, as mentioned at the beginning of this subsection.
We should pay attention to this group of students especially in \textit{Class 2}.
It seems that they could not fully grasp the contents specified during the face-to-face lecturing time and were always struggling to adapt themselves to our new style of teaching.
To them, the group problem solving might be a heavy burden because they felt uneasy with even basic concepts implemented in the problem setting.
However, the same group of \textit{Class 1} did not show the similar pattern of preference to the group activity even though they might feel that more reduction of the face-to-face lecturing time made them uneasy than in \textit{Class 2}.
We think that it was because GPS was designed to be optional in \textit{Class 1}
and students could decide whether they can join GPS or study the topics by themselves.
This explains why the satisfaction level with GPS for the group with $ -10 < \operatorname{FR-M} < 0$ of \textit{Class 2} is distinguishably lower than that of \textit{Class 1}.
Therefore, we specify these students in \textit{Class 2} as lecture-oriented students that need serious attention in a redesigned class based on flipped learning.

We also need to pay attention to the groups with $\operatorname{FR-M} < -10$ for further developing the education models based on flipped learning.
In the groups with $ \operatorname{FR-M} < -10$, there are 40 students from \textit{Class 1} and 43 from \textit{Class 2}.
The average scores of those students are reduced from 65.6 ($\pm$ 2.23, the standard error)  to 36.4 ($\pm$ 2.71, the standard error) for \textit{Class 1} and from 65.5 ($\pm$ 2.13, the standard error)  to 36.3  ($\pm$ 2.59, the standard error)  for \textit{Class 2}.
While their performances were higher than the average at the midterm exam but lower at the final exam,
their satisfaction level with L and especially GPS is as high as other $\operatorname{FR-M}$ groups.
In both classes, this group of students tended to evaluate our teaching method generously good as in Fig. \ref{fig:fr-m}.
When the course contents were easy to them before the midterm exam, they enjoyed the new class activities
As they learned new and challenging topics, they still had good impressions on the instructor's efforts but could not assess what they actually learned.

\section{Discussion}

In this paper, we have presented achievements and reflections of the students who have taken General Physics I,
the first part of the calculus-based introductory physics course for two semesters.
In Semester 2012, our primary concern was whether a combination of pre-class self-study based on IT technology and in-class group problem solving
could be a reasonable substitute for face-to-face lecture.
We have found that our redesigned course offered students a more effective way of learning physics than the traditional type of classes based only on lectures.
In Semester 2013, we controlled the proportion of the face-to-face lecture systematically over two classes,
while the remaining class hours were devoted to group problem solving.
In addition, students in the two classes were required to do pre-class self-study.
The students' achievements were not distinguishable in the sense that their statistical distributions of scores in the midterm and final exams were not different.

The fact that students in our redesigned class performed better than those in the traditional type of class in Semester 2012
suggests that the pre-class self-study and in-class group problem solving, the two features that are present
only in the redesigned class, play an important role in students' learning.
On the other hand, face-to-face lecture does not seem to figure predominantly in students' learning of physics,
because the students in the redesigned class performed better even though the lecture hours for them were
reduced to only 1/2 of those for the students in the traditional type of class.
Nevertheless, examination of data from Semester 2013 suggests that it is not desirable to completely eliminate
face-to-face lecture.
The data indicate that  the students in \textit{Class 2}
where the in-class hours are divided equally into face-to-face lecture and group problem solving have
performed as good as those in \textit{Class 1} where face-to-face lecture hours are further reduced.
One can say that the further reduction of lecture hours from 1/2 the entire in-class hours in
favor of group problem solving did not improve students' performance.
Our main conclusion therefore is that students learn more from pre-class self-study and peer instruction through
group problem solving than from traditional face-to-face lecture by an instructor,
and yet face-to-face lecture cannot be totally ignored.
While pre-class self-study should be retained as an effective learning strategy, it is desirable to assign
an appropriate proportion of the in-class hours to face-to-face lecture.

We need to mention the contrasts of the two classes in Semester 2013,
whose proportions of the face-to-face lecture were
reduced to 1/3 and 1/2, repectively.
First, a group of the students of the class where the lecture was reduced to 1/3,
who generally scored higher in the midterm and final exam,
tended to be satisfied with our new method of reducing
the proportion of the lecture and having enough time for group problem solving/discussion.
Second, a group of the students of the class where the lecture was reduced to 1/2,
whose final scores were slightly reduced than the midterm scores,
think that the storytelling lecture was rather useful for their learning than group problem solving activity.
This seems to suggest that these students feel uneasy about drastic reduction of face-to-face lecture hours from 1/2
those of the traditional type of class.

There are a few limitations in this research.
First, having considered that the second class of each week for \textit{Class 1} in Semester 2013 was optional, we could not study how the fact that students had more freedom in class schedule affect students' achievements.
The students in \textit{Class 1} had flexibility in class time schedule - they could choose to study the topics by themselves instead of coming to the optional problem solving classes.
Hence, they might have developed their time-management skills for study which could positively affect their achievement as \citet{bergmann1} pointed out.
The options given to the students of \textit{Class 1} might positively affect their achievements so that the results of the midterm and final exams of both classes were not different.
At least, our study seems to suggest that it would be safe to give students more options when the proportion of the face-to-face lecture is reduced in flipped classrooms, as we did in \textit{Class 1}.
Second, since our results are only from two semesters,
we should continue to test our model for longer terms with more students.
Even though our strategy of increasing the interactions among students and instructors
by reducing lecture hours was proven to be fruitful for the learning of the majority of students,
still there were students who could find benefits from the face-to-face lecture.
Whether our model is effective generically for all physics courses is still questionable.
Our experiments were done only on the first part of an introductory university physics,
and most freshmen have already been exposed to the topics in their high schools.
We are going to build up a similar model for the second part and study it using the similar method.

\begin{acknowledgements}

We thank Prof. Kook Joe Shin and Geunyoung Jin of Division of General Studies at UNIST,
and Prof. Georg Rieger, Prof. Sarah Gilbert and Prof. Carl Wieman of Carl Wieman Science Education Initiative  at the University of British Columbia
for many helps during the development
stage of our redesigned General Physics course.
We thanks Dr. Sungwook E. Hong at KIAS, Prof. Jeong-Ah Lee and  Prof. Doo Seok Lee at DGIST, and Dr. Min Kyu Kim at Ohio State University  for helpful discussion.

\end{acknowledgements}

\end{document}